# Hierarchical self-organization of highly-ordered granular ensemble of optical solitons through collective motions

Xiaocong Wang[1,2,†], Benhai Wang[1,2,†], Haochen Lin[1,2,3,†], Wenbin He[1,2,*], Yu Jiang[1,2],  Qi Huang[1,2], Xintong Zhang[1,2], Long Zhang[1,2,*], and Meng Pang[1,2,3*]

[1]*Russell Centre for Advanced Lightwave Science, Shanghai Institute of Optics and Fine Mechanics and Hangzhou Institute of Optics and Fine Mechanics, Hangzhou 311421, China*
[2]*Zhejiang Key Laboratory of Microstructured Specialty Optical Fiber, Hangzhou 31142l, China*
[3]*School of Physics and Optoelectronic Engineering, Hangzhou Institute for Advanced Study, University of Chinese Academy of Sciences, Hangzhou 310024, Zhejiang, China*

*Corresponding author:
Wenbin He, Email: wenbin.he@r-cals.com
Long Zhang, Email: lzhang@siom.ac.cn
Meng Pang, Email: pangmeng@siom.ac.cn

†These authors contribute equally to the work

## Abstract

Self-organizations of ordered patterns in far-from-equilibrium many-body systems host fundamental importance in many disciplines. Meanwhile, complex systems often feature hierarchical structures with distinct scales for different layers, enabling high-level effective dynamics without exhaustive tracking of all possible degrees of freedoms. In this work, we report a study of the self-organization dynamics of highly-ordered soliton ensembles in a high-harmonic mode-locked fiber lasers through collective motions driven by nonlocal optomechanical interactions and local collisions, which exhibit a series of universal characteristics reminiscent of phase transitions. Moreover, the multi-soliton laser-field can be coarsely grained as a granular ensemble of limit-cycle oscillators with simple interaction rules derived from fine-scale physics. The self-organization of the multitude of solitons in the mode-locked laser cavity can then be mapped into a low-dimensional dynamic model that essentially reproduced the emergent process. Our work affords a conceptual framework for understanding the complex structure formation in nonlinear laser systems, and may help to design ultrafast lasers by exploiting universal principles of collective motions.

## 1. Introduction

Complex many-body systems in far-from-equilibrium states often exhibit ubiquitous self-organization phenomena, leading to a variety of patterns [1, 2] or coordinated behaviours [3, 4] that are of fundamental importance in physics [5, 6], biology [2, 7] and engineering [8, 9]. Meanwhile, the ubiquitous self-organization phenomena have aroused great interest in search for universal governing rules irrelevant to their physical instantiations[2, 10]. However, it is generally difficult to determine the detailed dynamics of the vast degrees of freedom of complex systems, while the governing rules may remain obscured even if the details of every possible degree of freedom have been unfolded. Fortunately, complex systems in nature often feature layered structure of distinct scales in time or space and can thus be understood in a hierarchical manner using effective dynamics derived from coarsely-grained physics. In practice, self-organizations in a complex system can usually understood by decomposing the system into many subsystems with couplings, resembling interacting nodes in a network [11]. Even though each subsystem may already embody intricated structures and dynamics, the collective motions of the coupled subsystems can be understood at organizational levels without specific dependence on fine-scale details. Critical for such hierarchical analysis is the identification of essential collective variables (order parameters) that can best characterize the macroscopic patterns while enabling a simple and self-consistent dynamic theory with prominent reduction of degrees of freedom.

The self-synchronization of interacting subsystems with rhythmic motions [12, 13] has long been regarded as paradigm for self-organizations, which can produce coherent spatial or temporal patterns in macroscopic scales. Well-known examples are particularly abundant in many life processes, including the coordinated motion of cardiac pacemaker cells and firing neurons, synchronized flashing of fireflies and chorusing of crickets [14]. Self-sustained limit-cycle oscillators with stable amplitudes and free phases have proved to be a particularly fruitful model for modelling rhythmic subsystems in a many-body system [2]. In particular, such oscillators can cooperate to achieve a global rhythm through weak interactions which can overcome the differences in their intrinsic frequencies and noise perturbations. In addition, such self-synchronization phenomena often closely resembles phase transitions, featuring, symmetry breaking, emergent order-parameter dynamics, critical fluctuations, and hysteresis. [15].

Optical solitons in a mode-locked laser [16-18] represents a generic type of coherent optical structure emerging out of nonlinear coupling of many cavity modes [19, 20]. Hierarchically, these particle-like laser solitons can behave as fundamental building-blocks and form macroscopic temporal patterns [17, 21], given appropriate nonlinear couplings can be established between them. However, the self-organization of a multitude of laser solitons placed difficulties for both experimental manipulation and theoretical modelling [22]. In one hand, each laser soliton already represents a highly active subsystem with high-dimensional dynamics, while the nonlinear interactions between them needs to be carefully tailored in order to realize reconfigurable self-organized patterns. In the other hand, theoretical modelling of ultrafast laser often entails nonlinear differential equations with infinite spatiotemporal dimensions, leading to formidable computation complexity in case of macroscopic pattern scales. As a result, low-dimensional effective dynamics [2] derived through hierarchical analysis are demanded to establish an accessible patten-level understanding without unfolding all degrees of freedoms of the laser field.

In this work, we report the experimental characterization and dynamic modelling for the self-organization process of in a high-harmonic mode-locked soliton fiber laser. Spontaneous self-ordering of a multitude of laser solitons towards a stable temporal pattern can be reproducibly observed during the laser self-starting process, which is driven by nonlocal optomechanical interactions in a solid-core photonic crystal fiber (PCF) as well as local collisions between the solitons. While multifarious dynamic details can be observed during the self-organization process, the temporal patterns remain largely insensitive to these details while exhibiting universal properties reminiscent of phase transitions [12, 19]. Moreover, the distinct temporal scales involved in both the multi-pulse structure and pulse-interaction dynamics have enabled a low-dimensional effective model for a granular ensemble of interacting particles. The effective model has essentially captured the macroscopic feature of the self-organization process while highly resembling the Kuramoto self-synchronization transition. This work can afford a useful conceptual framework for understanding the complex structure formation in nonlinear laser systems and also provides a flexible instantiation to explore collective motions in nonlinear optical systems.

## 2. Illustration of Concept

A high-harmonic mode-locked cavity hosting a multitude of laser solitons provides a simple platform for generating on-demand multi-pulse laser-fields [23-26]. However, laser solitons generally suffer from random temporal "kicks" in the cavity due to amplified spontaneous emissions in the gain sections [27] and uncontrolled interactions[28, 29], leading to random drifting and eventually disordered temporal pattern. By introducing appropriate nonlinear interactions between the laser solitons, these random fluctuations can be overcome, leading to the generation of a regular macroscopic temporal pattern (see Fig.1a). In practice, while each soliton has a ps or sub-ps duration, the full laser-field could occupy the entire cavity with a span of tens or hundreds of ns (Fig.1b). The distinct structural scales enabled a granular model for the temporal soliton pattern in which each soliton can be regarded as a particle flowing on a ring in a periodic manner, while its fine-scale soliton envelope become irrelevant to macroscopic-level descriptions. As a result, the solitons are only defined by their positions and velocities on the ring, resembling the phase and frequency of limit-cycle oscillators.

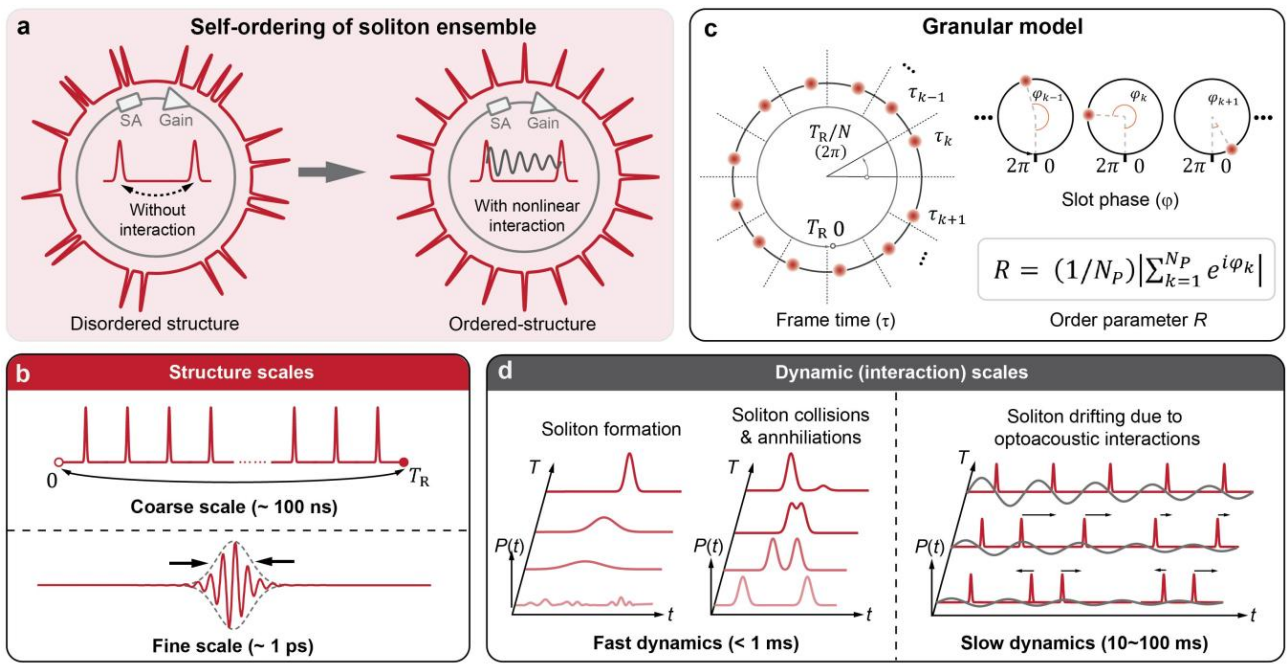

**Fig. 1 | Structures and dynamic model for self-organized laser-soliton ensemble. a.** The macroscopic structure of the multi-soliton ensemble in mode-locked laser cavity can spontaneously transit from disorder to order given nonlinear interactions between individual solitons. **b.** Distinct structural scales. While each soliton takes sub-ps duration, the entire ensemble spans over the entire cavity with ~100 ns scale. **c.** The granular model for the soliton ensemble in which solitons are modelled as particles following on a ring with limit-cycle oscillations. The position of each soliton is defined using frame time $\tau$ in each round-trip and slot-phase $\varphi$ in each time slot. Then an order parameter R can be defined to represent the macroscopic coherence of the temporal pattern. **d.** Distinct time-scales of different soliton dynamics. While soliton formations and collisions (which can lead to soliton annihilations) occurred at fast time scales (< 1 ms), the soliton drifting due to nonlocal

optomechanical interactions occurs at much slower time scales (10 – 100 ms). The sinusoidal curve beneath the solitons represents the optomechanical modulations driven by the solitons themselves as the source of nonlocal interactions.

An order parameter can be introduced to describe the patten-level coherence of the solitons in the laser cavity by transforming their temporal order into a collective phase relation, resembling the description for synchronizations of coupled oscillators [30, 31] (see Fig. 1c). The temporal order of each soliton is described by its "frame time", which is defined as its relative time $\tau$ in each round-trip (frame) with a span of $T_R$. Each frame is furtherly divided into a temporal grid with $N$ time-slots ($N$ equals to the target harmonic order [32]). The solitons located in its respective time-slot can be assigned with a "slot phase" ($\varphi$) by regarding the span of each slot as one phase cycle ($2\pi$). The order parameter $R$ is then defined as:

$$R = \frac{1}{N_p} \left| \sum_{k=1}^{N_p} e^{i\varphi_k} \right| \tag{1}$$

in which $N_p$ is the pulse number in the cavity. Note that the number of pulses $N_p$ (both during and after the self-organization) may not equal to the number of time-slots $N$. Meanwhile, an indefinite number of solitons could occupy a single time-slot during the process. When the soliton ensemble takes a regular lattice pattern at $N^{th}$ harmonic of the mode-locked cavity, $R$ reaches the highest value of unity even in the presence of empty time-slots. (See Supplementary Materials Section 1.)

In the dynamic aspect, both fast and slow dynamics can occur during the self-organization process, which correspond different interaction mechanisms and strengths. While fast formation of individual solitons and direct collisions between the solitons typically occur at a fast time scale (typically less than 1 ms), the drifting of the solitons due to nonlocal optomechanical interactions occur at such slower time scale (typically 10 to 100 ms, see Fig. 1d). In the light of the distinct time scales in dynamics, the soliton interactions in the granular ensemble can be essentially captured using a highly simplified functions. While the fast dynamics can be regarded as lumped and local actions with pre-defined outcomes, the slow drifting of the solitons due to optomechanical interactions is modelled using a sinusoidal coupling function derived from the physical mechanism. The simplified dynamics based on the granular model essentially reproduce the complex self-organization process observed in experiments, as we will see in the following sections.

## 3. Experiments

We built an optomechanically mode-locked soliton fiber laser based on a solid-core PCF [23, 24, 33, 34] as the platform to investigate the self-organization process of highly-ordered soliton ensemble. The PCF inserted in the mode-locked cavity has a core-resonance at 1.905 GHz, which enables long-range optomechanical interactions between the laser solitons [35] and eventually, a self-stabilized high-harmonic operation at 1.90 GHz ($310^{th}$ order) in the 34-m-long ring-cavity fiber laser [36]. (See setup details in Supplementary Materials Section 2.) In experiments, we switched on and off the laser gain of mode-locked cavity for many times and recorded the self-organization process of the intra-cavity laser fields (from noise-like background to highly-ordered soliton-ensemble formation). We found that the recorded processes all exhibited similar macroscopic features while having stochastic details in both fine and macroscopic scales. A typical self-organization process of the ultrafast laser-field after turning on the laser gain is shown in Fig. 2a. The intra-cavity light-field evolution was recorded and plotted in a moving frame with a span of ~163 ns (the steady-state round-trip time). Note that only about one-third of the solitons of the ensemble were plot due to the size limit. Initially, hundreds of laser solitons quickly emerged at random positions out of the noise-like background at few-ms scale after the laser gain is switched on, which can be regarded as the self-organization of lower-level coherent structures (see zoom-in plot in Fig.2c) [19, 37]. These newly emerged solitons then went through self-ordering process through collective motions due to optomechanical interactions. Inevitably, soliton collisions would occur due to attractive optomechanical interactions [25]. Despite the stochastic appearance of these collision dynamics, the outcomes of these soliton collisions were found to follow clear statistical rules. As shown by the typical collision event in Fig. 1d (left panel), all collisions in Fig.1a generally ended up with the survival of only one soliton after a brief interaction time of ~ 1ms [26]. Moreover, we found that the annihilation of solitons during these collisions was largely compensated by newly-emerged solitons at other positions inside the laser cavity (see a typical event in Fig.1d, right panel). The number of participating solitons exhibited a rapid growth at the beginning and then gradually settled down to a nearly constant value after a weak overshooting, which was attributed to the slow relaxation oscillation of the Er-doped gain fiber (see Fig. 2b). The pulse-number evolution presented in Fig. 2b indicates that a dynamical balance existed between the annihilation and regeneration of laser solitons throughout the entire self-organization process.

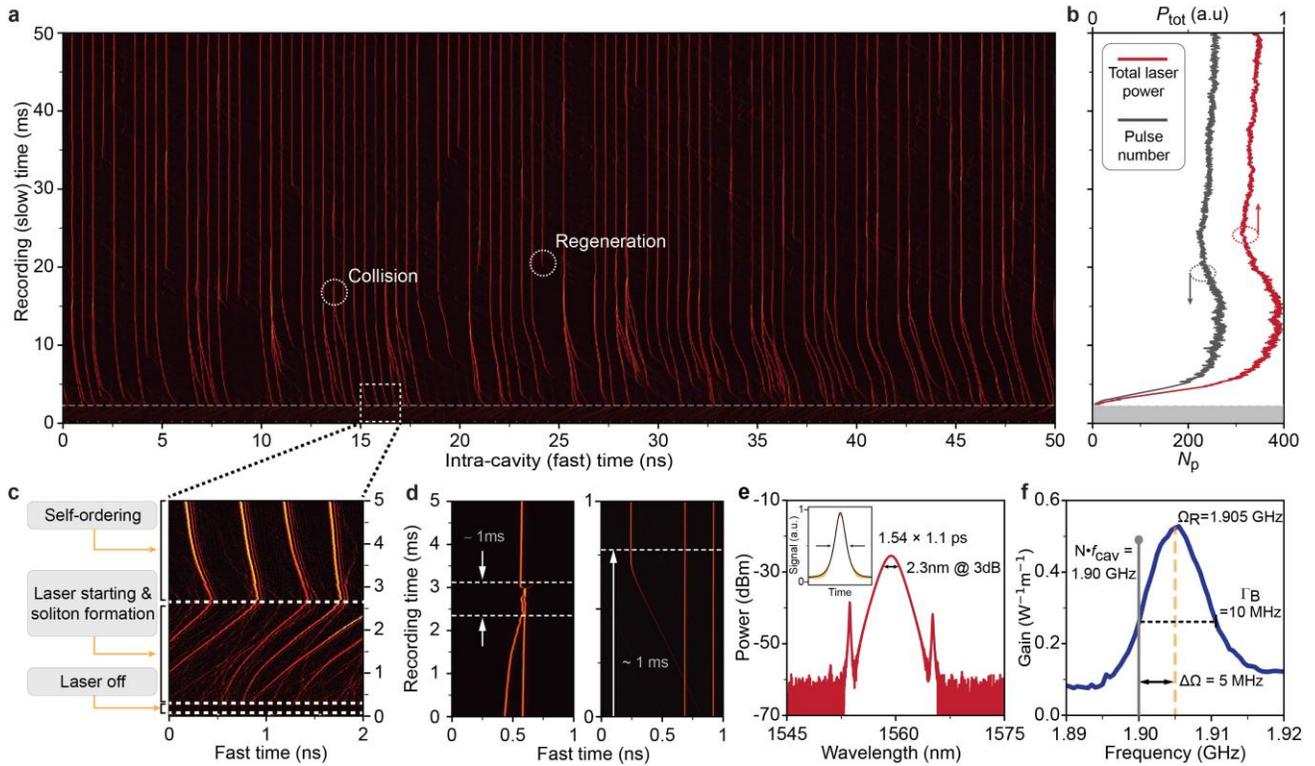

**Fig. 2 | Experimental Results. a.** Experimental recording of self-organization process of the soliton ensemble in the optomechanically mode-locked laser over 50-ms time. 100 out of the total 310 time-slots are shown. **b.** Evolution of the total pulse number and pulse energies. **c.** Zoom-in plot for the dashed-box part in a, which include the laser starting and soliton formation process after the laser gain if switched on, followed by self-ordering of these newly emerged solitons. **d.** Typical soliton collision (left) and regeneration (right) process during the self-organization process. **e.** Laser spectrum at steady state after the self-organization. Inset: Autocorrelation trace of the output solitons with hyperbolic-secant fitting. **f.** The acoustic gain spectrum (yellow) of the PCF. The steady-state repetition rate at locked at 1.9 GHz as indicated by the grey line, which is detuned from the resonance frequency by 5 MHz.

A highly-ordered lattice pattern of laser solitons is established typically after a few hundreds of milliseconds, though possibly with some empty time-slots. The lattice pattern represents a deterministic breaking of the translational symmetry of within the ensemble since the solitons could have taken arbitrary spacings due to random perturbations from noise [27]. The optical spectrum in the final stable state is show in Fig. 2e, featuring soliton-regime laser spectrum [38], while the pulse repetition rate is locked to the acoustic resonance of the PCF (see Fig.2f). Importantly, given fixed cavity length, the harmonic order (time-slot number) $N$ of the final states is fixed by the PCF acoustic resonance, despite the stochastic self-organization processes [32]. The positions of empty time-slots in the final pattern varied between different self-organization processes while their total amount depends on the pump power, which indicates the existence of multistability for the self-organized pattern, being a generic feature of nonlinear dissipative system (see Section 6 below).

Using the order parameter $R$, we can describe the pattern-level self-ordering dynamics of the soliton ensemble with significantly reduced degrees of freedom. As the self-organization process exhibit prominent stochasticity under the same cavity conditions, the evolution of $R$ also varied randomly from case to case, as shown in Fig.2a inset. However, these different evolutions have similar features, in which their order parameter $R$ grew up rapidly from trivial values after the laser gain is switched on (possibly with some oscillations), and then asymptotically reached the value of unity. More importantly, the averaged evolution of $R$ for multiple events was found to exhibit a prominent dependence upon the accumulated interaction strength between the solitons. By changing the detuning of the target cavity harmonics ($310^{\text{th}}$) from the acoustic resonance ($\Delta\Omega = \Omega_{\text{R}} - Nf_{\text{cav}}$) from 5 MHz to 9 MHz, we have noticed a prominent slowing-down of the build-up of $R$ (see Fig.3a).

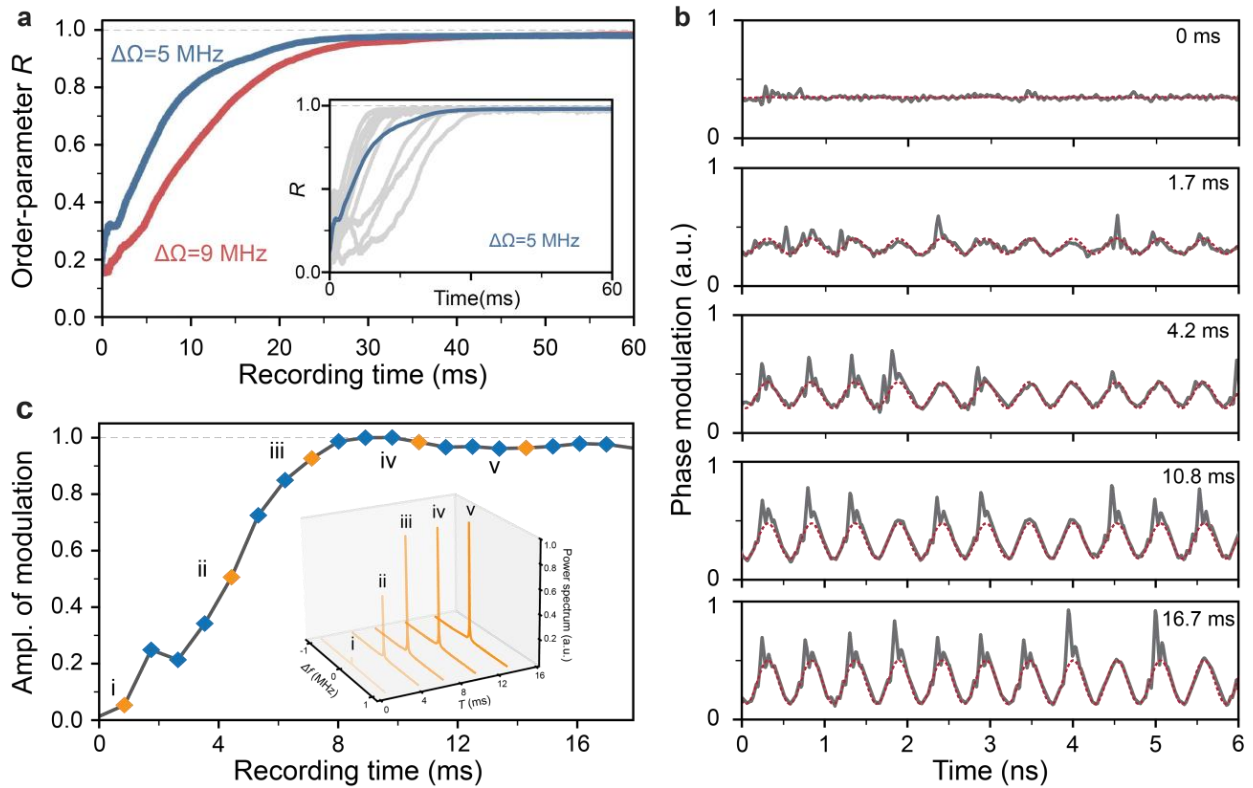

**Fig. 3 | Order parameter dynamics. a.** The averaged order parameter evolutions at different cavity resonance detuning of $\Delta\Omega$ (red: $\Delta\Omega$=9MHz; blue $\Delta\Omega$ = 5 MHz). Inset: Evolutions of $R$ for multiple self-organization process under the same condition given $\Delta\Omega$ = 5 MHz. **b.** Phase-modulation along the PCF measured during the self-starting process using homodyne technique. The phase modulation exhibits a sinusoidal component resulting from the optomechanical interactions. **c.** The evolution of the amplitude of the sinusoidal optomechanical modulation during the self-starting process, resembling that of order parameters in b. Inset: evolution of the power spectrum of the measured sinusoidal phase modulation during the self-organization process, featuring a growing peak at the final repetition rate.

Apart from the order parameter $R$ that concerns the coherence of the granular ensemble pattern, the self-organization process can also be indicated by collective optomechanical interaction between the solitons, which is dominated by the total phase modulation induced by the pulse-driven optomechanical effects in the PCF [24, 35]. While incoherent optomechanical interactions tend to neutralized each other, coherent interactions enhance the total modulation and impose regularities within the pattern, which furtherly reinforced the modulation as a back-action. Using the homodyne measurement technique [36], we directly measured the real-time phase modulation along the PCF during the self-organization process, which shows a sinusoidal waveform accompanying the pulse sequence with a growing magnitude (see Fig. 3b). This sinusoidal phase modulation can be regarded as the consequence of the combined acoustic phase modulation driven by all the solitons, and the evolution of its amplitude resembles that of the order parameter $R$ (see Fig. 3c). The total optomechanical modulation can also serve as an order parameter for the soliton ensemble with a dual-role characteristic: it indicates the coherence of the soliton pattern which determines its driving strength upon the acoustic resonance, while as it also imposes order to the soliton pattern by providing coherent phase modulations.

## 4. Low-dimensional dynamical model

The self-organization of the highly-ordered soliton ensemble can be described using a granular model with low-dimensional dynamic system such that this entire process can be regarded as a self-ordering process of a many-particle system following simple nonlinear interaction rules. Each soliton in the cavity, being a subsystem of entire structure, is regarded as a single particle flowing periodically a ring such that merely the positions and velocities of the subsystems are concerned in the model (see Fig. 4a, middle panel). The interactions between these solitons, either through optomechanical interactions or direct collisions, can be simplified as coupling functions of different topologies and strengths that determines the position and velocity changes of each particle after each round-trip. All the laser solitons are set to have the same intrinsic round-trip frequency $f_0$. This assumption is generally valid since the pulse energy and central frequency for all the solitons are largely fixed by identical gain/loss balances in the mode-locked laser [39], although they are inevitably exposed to noise perturbation that needs to be robustly cancelled out by the nonlinear interactions between them.

Given an overdamped limit for soliton motion in the cavity [36], an one-dimensional dynamic can be established for each particle in the granular ensemble in form of an iteration map (at a time interval of $1/f_0$) as following:

$$T_R \frac{d\tau_k}{dT} = \Phi_{k,\tau} - \sum_{j=1}^{N} F_A(\tau_k, \tau_j) - D_\tau(\tau_k, \tau_{j,C}) \qquad (2)$$

in which, $\tau_k$ is the frame time of the $k^{th}$ particle, $\Phi_{k,\tau}$ the noise added upon the frame time of the particle after each iteration. $F_A(\tau_k, \tau_j)$ denotes the optomechanical interaction between each pair of particles, which can be described using Green's function [40] with a sinusoidal dependence on their frame time difference $\Delta\tau_{kj}$, i.e.

$$F_A(\tau_k, \tau_j) = iC_0Ae^{\left(-\Gamma_B + i\sqrt{\Omega_R^2 - \Gamma_B^2}\right)\Delta\tau_{kj}} \tag{3}$$

in which $\Omega_R$ is the resonance frequency of the PCF-core, $\Gamma_B$ the acoustic linewidth, $A$ the strength of the electrostrictive force, which we assume to be identical for all particles, and $C_0$ is a constant. Each soliton excites an optomechanical phase modulation along the PCF, which can be experienced by the following solitons, while each soliton also experiences the overall phase modulation excited by all the precedent solitons (see Fig. 4a, left panel). Note that $\Delta\tau_{kj}$ needs to be calculated in a unidirectional fashion (See Supplementary Materials Section 1). $D_\tau(\tau_k, \tau_{j,C})$ denotes the coupling functions due to local direct collisions, in which $\tau_{j,C}$ denote the frame time of the soliton that collide with the $k^{th}$ particle. Due to the brief interaction time and deterministic outcomes, a simplified collision rule is adopted for the granular model as below: When two neighboring particles reach below a threshold spacing, the former one gets a random kick by its position, while the other disappears and emerges at a random position in the cavity (see Fig. 4a, right panel). This simple rule can capture the essence of the collision dynamics from a macroscopic view despite the stochastic transient details. The LHS of Eq. (2) represents the discrete frame-time change of each particle after each iteration (i.e. the "velocity"). In steady-state condition, this term should become a constant value (could be non-zero) being identical for all the solitons. Apart from the collision terms, the dynamic system describe in Eq. (1) highly resembles the well-known Kuramoto model [41, 42] that describes coupled oscillators with sinusoidal coupling functions dependence on their phase differences (see Supplementary Materials Section 3). The mathematical resemblance between these two dynamic models naturally leads to analogous results in terms of self-synchronization for rhythmic motions, indicating the fundamental principle that similar formulations beget similar physics.

Using the dynamic model described as Eq. (5), we performed a numerical simulation for a many-particle system with 100 interacting particles flowing on a ring with a frame span of 100 ns and a time-slot span of 1 ns. The acoustic resonance frequency is set to be 1.005 GHz with a bandwidth of 10 MHz, which is detuned away from the $100^{th}$ harmonic of cavity round-trip frequency by 5 MHz [24]. The particles are randomly positioned in the ring as the initial condition in accordance with experimental observations shown in Fig. 2c. The resultant simulation exhibits an emergent self-ordering towards a regular temporal pattern over 5000 iterations despite the initial randomness and noise perturbations. All the essential details we observed in experiments have been reproduced in the simulation, namely the slow drifting, the collisions, and the regeneration of particles (see Fig. 4b). The corresponding evolution of order parameter $R$ also exhibit varied trajectories for different simulation cases given the same coupling functions and random initial condition. (See Fig.4c). The averaged evolution, nevertheless, also exhibits a reproducible dependence on the accumulated interaction strength in accordance with experimental results in Fig.3a, which can be seen by the two evolution curves in Fig. 4d corresponding to different frequency-detuning $\Delta\Omega$ of the cavity harmonic from the acoustic resonance.

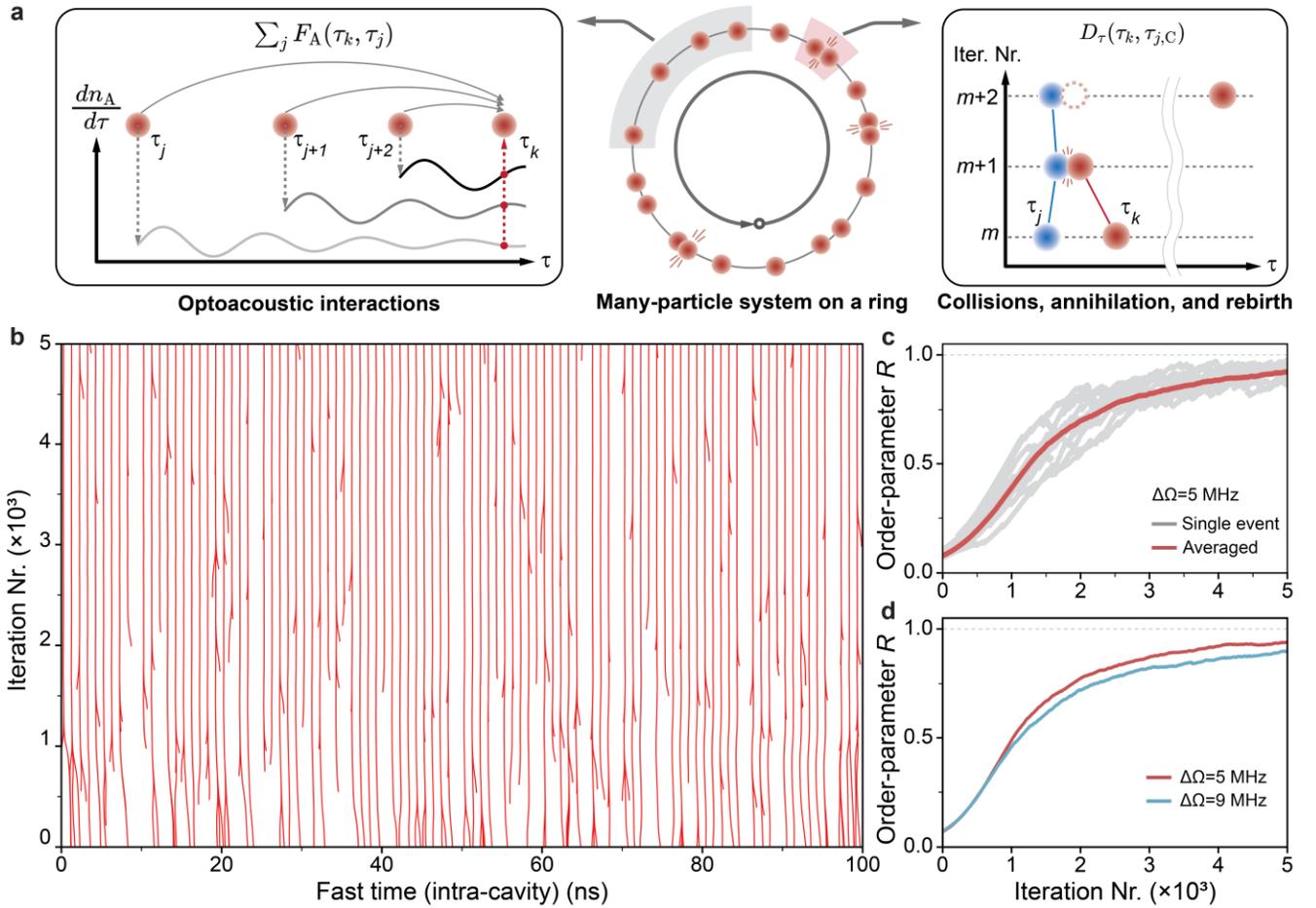

**Fig. 4 | Low-dimensional model and simulation results. a.** Concept of the model based on a simplified system of many-particle flowing on a ring. The particles have long-range optomechanical interactions described by summed sinusoidal coupling function $\sum_j F_A(\tau_k, \tau_j)$ (left panel) as well as short-range direct collisions described by lumped-action function $D_\tau(\tau_k)$ (right panel). **b.** Typical simulated evolution of self-ordering process for 100 particles over 5000 iterations. The time-slot span is 1 ns while the acoustic resonance is 1.005 GHz. **c.** Evolutions of order parameter $R$ for different simulation events under the same coupling functions and random initial condition. **d.** Averaged order parameter evolution based on simulated self-organization process given different detuning (5 MHz and 9 MHz) from the acoustic resonance.

## 5. Self-healing and ensemble-level entrainment

The self-organized, highly-ordered soliton ensemble in the mode-locked cavity hosts prominent self-healing ability given abrupt external perturbations. In the experiment, we launched a random sequence of external pulses into the cavity such that all the solitons are perturbed simultaneously and randomly in both their temporal positions and pulse energies [36], leading to a temporarily reduced order parameter (see Fig. 5a). We observed that the solitons have managed to quickly restore their temporal order and also pulse energies over a few milliseconds. Using the slot phase (defined as Fig. 3a) and pulse energy as the two variables for describing each soliton, the self-healing process can be presented in a polar coordinate. The states of these solitons are represented by a collection of state points in this coordinate, which temporarily dispersed and then quickly reconcentrated, as illustrated in Fig. 5b.

Importantly, a global net-shift of the averaged slot-phase of the soliton ensemble can be noticed after the self-healing process (see the first and the last frame in Fig. 5b), which probably resulted from a non-trivial net entrainment upon the ensemble induced by the external pulses. In fact, the self-organized soliton ensemble can also be simply entrained by pump modulations which can temporarily vary the group velocity (round-trip time) of all the laser solitons. As illustrated in Fig. 5c, the entire ensemble was adiabatically entrained during pump-power modulations while maintaining their coherence. This shift indicates that the entire soliton ensemble can be freely entrained as a single unit by external modulations without affecting the temporal pattern inside the ensemble. Therefore, we can safely postulate that the structural hierarchy can be further extended to higher levels using many of such ensembles as subsystems and establishing strong couplings between them. This may possibly enable the formation of self-synchronized laser array [43] that could deliver more-complex optical structures.

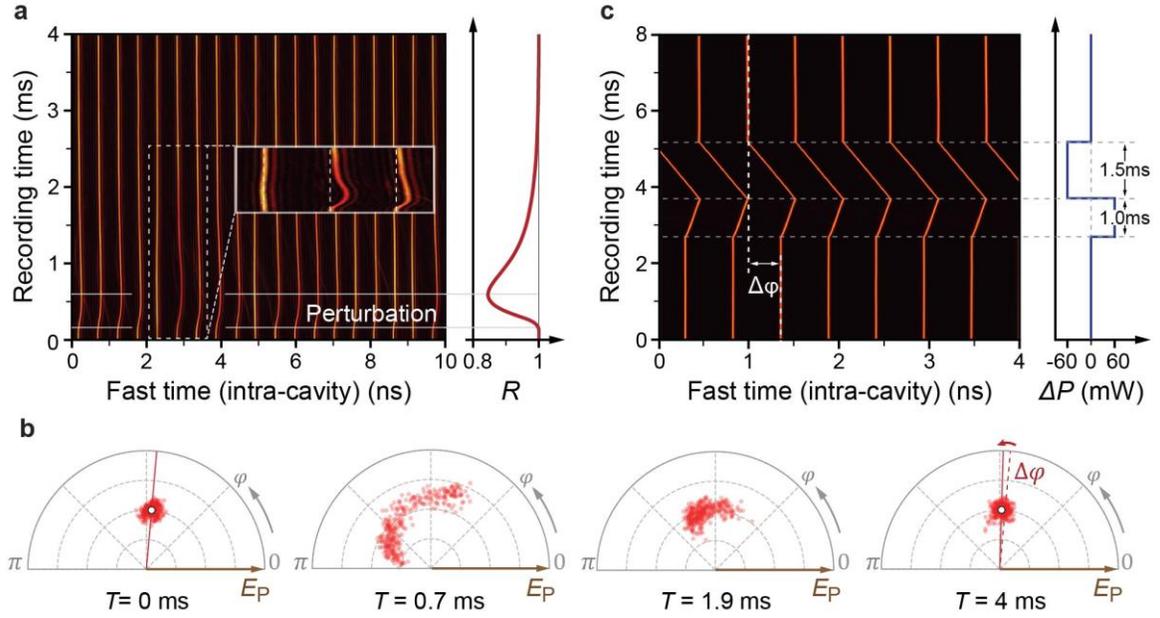

**Fig. 5 | Self-healing and ensemble-level entrainment. a.** Self-healing process of the soliton ensemble recorded in the experiments (19 out of 310 time-slots are shown). The while-box marked the interval of external perturbation. The inset gives a zoomed-in plot for a few perturbed solitons in the dashed box. The corresponding evolution of order parameter is shown in the right panel. **b.** The self-healing process shown in polar coordinate. The red dots are the state points regarding the slot phase and pulse energy of each soliton. The white dot at frame time $T$ of 0 ms and 4 ms marked the centre of mass for all points, which exhibit a slight shift in the averaged slot phase ($\Delta\varphi \approx 0.08$ rad). **c.** Ensemble entrainment through pump power modulation, which varies the soliton group velocity and leads to a global shift of the slot phase for all the solitons after the modulation.

## 6. Multistability

The nonlinear nature of coupling functions between subsystems generally leads to multistability for self-organized structures, which corresponds to multiple attractors for the collective variables of the system [44]. In addition, the system may reach different attractors with different probabilities out of the self-organization process or noise perturbations [2]. For the self-organized soliton ensemble with nonlocal sinusoidal optomechanical couplings, multistability exists in form of a variety of possible lattice patterns with empty time-slots given a fixed harmonic order. In this case, the order-parameter $R$ defined in Eq. (1) no longer suffice an adequate collective variable. To better distinguish patterns with different amount of empty time--slots, we introduce the grid filling-ratio ($F$) as a new collective variable, which is defined as the ratio of filled slots given a fixed harmonic order (see Fig. 6a). We have revealed that the self-organized soliton ensemble tends to acquire an $F$ value that falls into a narrow distribution, while the mean value of the distribution may shift prominently with the cavity-parameter variations. The statistical results shown in Fig. 6b illustrate three well-separated distributions of $F$ under different pump-power levels, while the corresponding $R$ of these patterns have only trivial changes (all being close to unity). We expect that distinct patterns with the same $F$ value could be further identified using additional order parameters (e.g. binary encoding [24]), which would be helpful to unveil the fine and hidden structures of attractors and basins in the phase space of the nonlinear laser system.

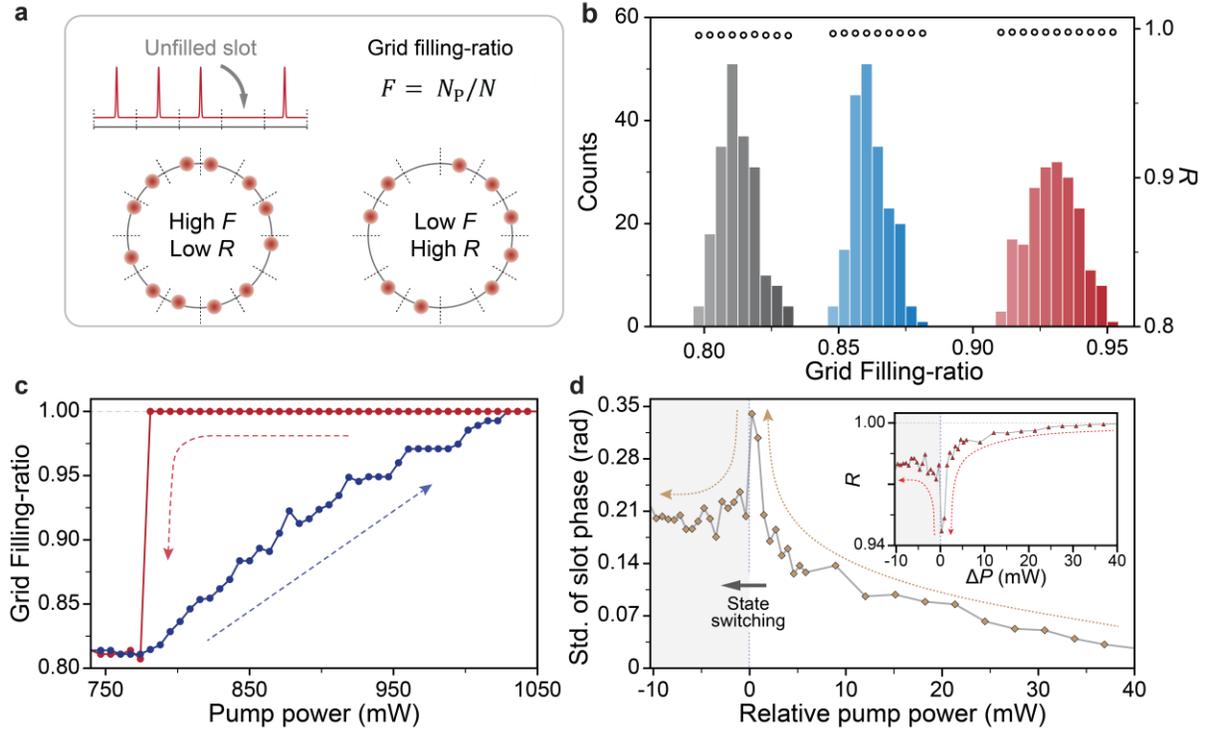

**Fig. 6 | a.** Definition of grid filling-ratio as new collective variable in addition to order parameter $R$. **b.** Distribution histograms of grid filling-ratio of measured self-organized soliton ensemble under different pump power (red: 660 mW, Blue: 625 mW, Black: 590 mW). Each histogram is retrieved from 200 self-organization events under the same control parameters. The correspondingly measured order parameters $R$ are marked as red dots for each different values of $F$. **c.** Hysteresis diagram of grid filling-ratio as the pump power is gradually varied between 450 mW and 900 mW. **d.** Critical fluctuation during state switching of soliton ensemble when gradually decreasing the pump power across the critical value denote as the origin point. Inset: critical change of order parameter $R$ during the pump power decrease.

In addition, we noticed that the self-organized soliton ensemble shows hysteresis effect when traversing different states as we varied the pump power, which can also be indicated by the grid filling-ratio $F$. When we slowly increased the pump power from 450 to 900 mW, $F$ gradually increased from 0.81 to 1.0. However, when the variation is reversed, the pattern is fixed ($F$ keeps the unity value) until a threshold beyond which $F$ abruptly decreased to 0.81 again (see Fig.6c). Moreover the soliton ensemble exhibit strong enhancement of slot-phase fluctuations when the system is close to state-switching point, reminiscent of the critical fluctuations phenomena in generic phase transitions [2]. A typical result is shown in Fig.3d, in which the system went through a state switching in terms of abrupt decline of grid filling-ratio as the pump power is decreased below the critical pump power. We can readily notice that the slot-phase fluctuation increased rapidly by nearly one order of magnitude before the state-switching and decreased suddenly to a lower value after the switching. The order parameter $R$ exhibits a similar critical change although the ensemble has maintained a high order parameter throughout the process. The multistability feature of the self-organized soliton structure concerns pulse energy distribution governed by gain and loss dynamics in the laser cavity, which cannot be analysed using the current granular model. Nevertheless, these features reminiscent of phase transitions should be of great interests for understanding the universality of collective motions in multi-pulse laser systems.

## 7. Discussions

During the self-organization process of the soliton ensemble, while the nonlocal optomechanical interactions have been well understood, the local collision-based interactions remain elusive, largely because the exact mechanisms possess intricate coupling between optical nonlinearity and dissipation concerning fine-scale soliton structures. Therefore, the coupling functions employed for collisions are phenomenological and probabilistic based on the experimental observations. Nevertheless, the low-dimensional model can still largely reproduce the pattern-level features despite the absence of fine-scale details, which highlights the critical isolation of effective physics for different layers in hierarchical structures. In addition, the collisions may have varied characteristics given different cavity parameters. For example, we observed in our experiments that the colliding solitons in some cases may experience strong repulsion without the annihilation of either solitons. In this case, the collision dynamics can still be captured using a modified rule, without diving deeply into the exact mechanism for the collision behavior (See Supplementary Materials Section 4).

The current model mainly focused on the self-ordering process of existing solitons, while a great number of dynamic details lies in the energy evolution of each soliton particularly during their formation and collisions, which may require the pulse energy as an additional degree of freedom for current granular model. We have recently added a pulse-energy dimension for analyzing the retiming dynamics of a single soliton in steady-state harmonic mode-locking [36]. However, for an ensemble of a multitude of solitons, the dynamics become sophisticated due to laser-gain competition between the solitons, which could lead to stochastic energy evolutions [45]. The inevitable soliton annihilation during the self-organization process may place great difficulties for analytical modelling, since the vanished solitons no longer have meaningful motions. Meanwhile, new solitons continuously emerged and added into the system, the mechanism of which has attracted extensive interest in recent several years [22, 37]. The current simplified model has circumvented such issues by regarding the annihilation and formation of solitons as an abrupt position shift of the soliton. A multi-scale modelling would probably be required to obtain comprehensive understanding of the nonlinear laser system, which consists both the particle-level description that concerns particle interactions as well as the cavity-mode-level description that concerns the pulse-energy dynamics [19, 20].

The self-organized soliton ensemble investigated in our experiments can have enhanced diversities in terms of the node types and network topology. The subsystems involved in the pattern formation may include both single solitons and various types of compact bound-state solitons as the interacting subsystems [11], each having distinct inner structures and dynamics. Moreover, the interactions between these subsystems can also be driven through different mechanisms such as dispersive-wave mediated [25] and gain-induced interactions [46]. In addition, the stability and the flexibility of the self-organized soliton ensemble could be further investigated through its multistability characteristics, which may help to gain insight of the dynamic phase portrait the nonlinear system with potential resemblance of other nonequilibrium phase transitions [12, 30].

## 8. Conclusions

In conclusion, we have experimentally and theoretically demonstrated the hierarchical self-organization dynamics of the massive-soliton ensemble in a mode-locked soliton fiber laser, revealing universal phase-transition-like characteristics such as emergent order-parameter dynamics, self-healing, multistability, hysteresis and critical fluctuations. By coarsely-graining individual solitons as particles flowing periodically on a ring with interaction rules derived from fine-scale dynamics, we established a low-dimensional model reminiscent of Kuramoto model, which faithfully reproduces the observed dynamics of ordering formation while dramatically reducing computational complexity. This hierarchical approach not only decouples pattern-level physics from intricate lower-level details, but also provides a unifying framework for understanding collective synchronization and structural formation in far-from-equilibrium many-body systems. Beyond ultrafast photonics, the conceptual and methodological advances presented here may help to design mode-locked lasers with better stabilities and customized performances [47] and ultrafast laser sources based on other platforms [48]. They may also inspire cross-disciplinary studies of manifold learning [49] and emergent ordering in complex networks [50, 51], as well as novel functionalities in areas such as laser-array control [43] and neuromorphic computing [52, 53], where collective dynamics and hierarchical structures play decisive roles.

## Methods

### Experimental set-up

The mode-locked fiber laser we built is a unidirectional soliton ring-fiber laser based on nonlinear polarization rotation (NPR). The laser cavity employs a Er-doped fiber as the gain section and a solid-core PCF to providing the optomechanical interactions. The total cavity length is ~ 34 m, which can host 310 time-slots in total at stable operation of 1.9 GHz. The cavity had a net anomalous dispersion of $-19$ $ps^2/km$ to ensure that the fibre laser operated in the soliton regime. The self-organization process in the mode-locked laser is initiated by the laser gain switching using external laser signal. The laser output sequence of pulse is detected using a high-speed photodetector and oscilloscope with a bandwidth of 59 GHz. To observe the self-healing dynamics of the self-organized soliton ensemble, we externally injected a sequence of randomly spaced pulses to perturb all the existing solitons with different strengths. The perturbation technique resembles the method given in Ref [36]. The random pulses were generated by modulation of CW signal through an electrical optical modulator (EOM) driven by an arbitrary wave generator (AWG), leading to a pulse duration of 70 ps and a peak power ~10 W after a EDFA section. The external pulses perturbed the intracavity solitons through cross-phase modulations within a 12-m-length of SMF in the cavity, while these external pulses are set to be blocked by the intra-cavity polarizer to avoid further perturbation of the laser gain. See details about experimental set-up in Supplementary Materials Section 2.

**Theoretical modelling**

We first consider the optomechanical coupling which occurs in the PCF in the cavity. For a fixed position in the PCF, the arrival of the $j^{\text{th}}$ soliton at the time $\tau_j$ causes an abrupt change in the time derivative of the refractive index of the PCF through electrostrictive force and is then experienced by the $k^{\text{th}}$ soliton arrived at this position time $\tau_k$, which can be expressed using Green's function.

$$\frac{dn_{k,j}}{dt} = \Theta(\tau_k - \tau_j)iAe^{\left(-\Gamma_{\text{B}} + i\sqrt{\Omega_{\text{R}}^2 - \Gamma_{\text{B}}^2}\right)(\tau_k - \tau_j)} \tag{4}$$

in which $\Theta$ is the Heaviside function, $\Omega_{\text{R}}$ the resonance frequency of the PCF-core, $\Gamma_{\text{B}}$ the acoustic linewidth, and $A$ the strength of the electrostrictive force, which we assume to be identical for all the solitons. Similarly, all the other $(N_{\text{p}} - 1)$ solitons ahead of the $k^{\text{th}}$ soliton would induce their own change in $dn/dt$, which can be summed up as $dn_k/dt = \sum_{j=1}^{N_{\text{p}}} dn_{k,j}/dt$. Then, a frequency-shift $\delta\omega_k$ is induced upon the $k^{\text{th}}$ soliton due to the optoacoustic index change ($\delta\omega(t) = -\partial\varphi_{NL}/\partial t$), which further leads to an incremental change of the group velocity, which can be described as

$$\delta\Delta v_k = -\beta_2 v_0^2 \delta\omega_k = \beta_2 v_0^2 k_0 L_{\text{P}} \frac{dn_k}{dt} \tag{5}$$

in which $k_0$ is the wavevector in vacuum, $v_0$ is the reference group velocity of the frame window, and $\Delta v_k$ is the relative velocity of the $k^{\text{th}}$ particle versus $v_0$. Then the resulting discrete acceleration of the $k^{\text{th}}$ soliton after one round-trip can be written as:

$$\delta\Delta v_{k,\text{A}} = C\sum_{j=1}^{N_{\text{p}}} iAe^{\left(-\Gamma_{\text{B}} + i\sqrt{\Omega_{\text{R}}^2 - \Gamma_{\text{B}}^2}\right)\Delta\tau_{kj}} = \sum_{j=1}^{N_{\text{p}}} F(\tau_k, \tau_j) \tag{6}$$

in which $C = \beta_2 v_0^2 k_0 L_{\text{P}}$ and $F(\tau_k, \tau_j) = iCAe^{\left(-\Gamma_{\text{B}} + i\sqrt{\Omega_{\text{R}}^2 - \Gamma_{\text{B}}^2}\right)\Delta\tau_{kj}}$. Note that the time interval $\Delta\tau_{kj}$ is specifically defined in a unidirectional way within each frame window. For solitons ahead of the $k^{\text{th}}$ soliton (i.e. arrived the PCF earlier than the $k^{\text{th}}$ soliton), we have $\Delta\tau_{kj} = \tau_k - \tau_j$, while for solitons behind it, we have $\Delta\tau_{kj} = \tau_k - \tau_j + T_{\text{R}}$, which actually account for the impact of solitons after the $k^{\text{th}}$ soliton in the previous frame window assuming a slow evolution speed. (See details in Supplementary Materials Section 1.) For $k = j$, we have $\Delta\tau_{kj} = 0$. In practice, there would be a trivial walk-off of two interacting solitons resulted from their different velocity over a single trip along the PCF, which has been neglected for within each iteration in simulations.

For the collision-induced coupling, the simplified collision rules indicate different consequence for the colliding solitons depending on their temporal order. For the $k^{\text{th}}$ soliton that collide with the $j^{\text{th}}$ soliton ahead of it (denoted with the additional subscript "C") given their spacing below a threshold value $\varepsilon_{\text{th}}$ ($0 < \tau_k - \tau_{j,\text{C}} < \varepsilon_{\text{th}}$), the $k^{\text{th}}$ soliton takes a new random position following a uniform distribution in the entire cavity, as well as a new relative velocity following a normal distribution. The coupling functions can be written as:

$$D_\tau(\tau_k, \tau_{j,C}) = -\tau_k + \eta_k, \qquad \eta_k \sim \mathbf{U}(0, T_R) \qquad (7)$$

$$D_v(\tau_k, \tau_{j,C}) = -\Delta v_k + \zeta_k, \qquad \zeta_k \sim \mathbf{N}(0, \sigma_v) \qquad (8)$$

in which $\tau_{j,C}$ and $\Delta v_{j,C}$ are the current frame time and relative velocity of the $j^{th}$ soliton that collide with the $k^{th}$ soliton, $T_R$ is the frame-window span, and $\sigma_v$ is the standard deviation for the new random velocity. The first terms on the RHS ($-\tau_k$ and $-\Delta v_k$) are present to erase its memory of previous states. Meanwhile, for the $k^{th}$ soliton to collide with $j^{th}$ soliton behind it ($0 < \tau_{j,C} - \tau_k < \varepsilon_{th}$), the position (frame time) of it gets a random kick, while the velocity acquired a random value close to zero. The coupling functions then become

$$D_\tau(\tau_k, \tau_{j,C}) = -\eta_k, \qquad \eta_k \sim \mathbf{U}(0, T_R/2N) \qquad (9)$$

$$D_v(\tau_k, \tau_{j,C}) = -\Delta v_k + \zeta_k, \qquad \zeta_k \sim \mathbf{N}(0, \sigma_v) \qquad (10)$$

For two colliding solitons, these two sets of coupling functions will be simultaneously fulfilled for each of the solitons respectively. In the cases where the $k^{th}$ soliton is intimately sandwiched by two other solitons (with both spacings below $\varepsilon_{th}$), the soliton in the middle is set to first collide with its nearest neighbor.

Using the coupling functions defined above, we can establish a two-dimensional dynamic model for each particle in the granular ensemble. We employed an iteration map for describing the discrete change of particle velocity $\Delta v_k$ and position (frame time $\tau_k$) of the $k^{th}$ particle in the ensemble over consecutive round-trips. The corresponding two-dimensional system can be expressed as:

$$T_R \frac{d\Delta v_k}{dT} = \sum_{j=1}^{N} F_A(\tau_k, \tau_j) + G(\Delta v_k) + D_v(\tau_k, \tau_{j,C}) + \Phi_{k,v} \qquad (11)$$

$$T_R \frac{d\tau_k}{dT} = -\frac{\Delta v_k}{v_0^2} + D_\tau(\tau_k, \tau_{j,C}) + \Phi_{k,\tau} \qquad (12)$$

in which $G(\Delta v_k)$ denotes the effect of gain filtering in the mode-locked laser, which provides an effective damping for the particle, i.e. $G(\Delta v_k) = -\gamma \Delta v_k$ [24]. Given the overdamped limit, which is in accordance with our previous observations [36], the two-dimensional system above can be simplified into a one-dimensional system as given in the Eq.(2), in which $F_A = -F/v_0^2\gamma$ and $C_0 = -\beta_2 k_0 L_P/\gamma$. Note that the impact of noise and collisions upon particle velocities has been neglected under the overdamped limit. The steady-state conditions for the two-dimensional model are that the discrete velocity change vanished to zero while the discrete frame-time change could be a non-zero constant value being identical for all the solitons. In practice we choose to induced a biased moving frame for plotting the self-organization process in which the final soliton ensemble remains fixed in the frame while the initial solitons may have a net global drifting. (See more details of the dynamic model in Supplementary Materials Section 3.)

**Data availability:** The data that support the plots within this paper and other finding of this study are available from the corresponding authors upon reasonable request.

**Code availability:** The numerical simulation code used in this paper is available from the corresponding author upon reasonable request.

**Funding information:** This work is supported in part by the National Natural Science Foundation of China (Grant No. 62375275 to Wenbin He and No.62275254 to Meng Pang), Strategic Priority Research Program of the Chinese Academy of Science (XDB0650000 to Long Zhang, Meng Pang, Zhiyuan Hung and Wenbin He), Shanghai Science and Technology Plan Project Funding (Grant No.23JC1410100 to Meng Pang and Wenbin He), and Fuyang High-level Talent Group Project (Wenbin He, Meng Pang, Xin Jiang and Philip Russell).

**Acknowledgement:** We thank Philip Russell for helpful discussions.

**Author contributions:** W.B.H. and M.P. conceived the work. X.C.W, B.H.W., H.C.L., Y.J., and Q.H. carried out the experiments. W.B.H. conceived the theoretical model. X.C.W. carried out the numerical simulations. X.C.W., B.H.W., H.C.L. W.B.H., X.T.Z, and M. P. made the theoretical and experimental analysis and wrote the manuscript. W.B.H., P. M. and L.Z provided fundings for the project. All authors contributed to the discussion of the results and the editing of the manuscript.

**Competing interests:** Authors declare no competing interests.

**Additional information:** Supplementary materials available

# Supplementary Materials for

# "Hierarchical self-organization of highly-ordered granular ensemble of optical solitons through collective motions"

Xiaocong Wang[1,2,†], Benhai Wang[1,2,†], Haochen Lin[1,2,3,†], Wenbin He[1,2,*], Yu Jiang[1,2], Qi Huang[1,2], Xintong Zhang[1,2], Long Zhang[1,2*], and Meng Pang[1,2,3*]

[1]*Russell Centre for Advanced Lightwave Science, Shanghai Institute of Optics and Fine Mechanics and Hangzhou Institute of Optics and Fine Mechanics, Hangzhou 311421, China*

[2]*Zhejiang Key Laboratory of Microstructured Specialty Optical Fiber, Hangzhou 31142l, China*

[3]*School of Physics and Optoelectronic Engineering, Hangzhou Institute for Advanced Study, University of Chinese Academy of Sciences, Hangzhou 310024, Zhejiang, China*

*Corresponding author:*
*Wenbin He, Email: wenbin.he@r-cals.com*
*Long Zhang, Email: lzhang@siom.ac.cn*
*Meng Pang, Email: pangmeng@siom.ac.cn*

[†]*These authors contribute equally to the work*

# Contents

# 1. Temporal order description of soliton ensemble dynamics

Highly-ordered soliton ensemble in a harmonically mode-locked soliton fiber laser is modelled as a granular ensemble of $N_{\mathrm{p}}$ moving particles in a unidirectional ring with narrowly distributed repetition rate and interactions, while exposed to weak noise perturbations and environmental fluctuations. These particles then become effectively a series of coupled limit-cycle oscillators (with a fixed amplitude) or simply a collective flow of particles on a ring. Given a steady-state repetition rate of $f_0$, the ensemble flowing on the ring can be described using a discrete iteration map, in which snapshots are taken at every $T_{\mathrm{R}} = 1/f_0$ interval. Particles with an instantaneous repetition rate of $f_0$ should remain fixed in the position on the ring over consecutive snapshots. Deviation of the repetition rate from $f_0$ (by $\Delta f_k = f_k - f_0$) as well as the noise perturbation $\xi_k$ upon the velocity will lead to accumulated shifts of particle positions. In principle, the dynamics of each particle $k$ can be described by their position in the cavity $p_k$. Without interactions, then incremental change of position along the map can be written as:

$$\delta p_k = T_{\mathrm{R}} \frac{dp_k}{dT} = \Delta f_k T_{\mathrm{R}} + X_k \tag{S1}$$

in which $X_k$ can be regarded as a random step following a normal distribution with limited standard deviation $\sigma_p$, i.e. $X_k \sim \mathbf{N}(0, \sigma_p)$. We added the subscript $k$ to indicate that each particle has independent random steps.

In practice, however, we prefer to describe the ensemble using its temporal order in the output from the laser cavity rather than the spatial position of each soliton in the cavity. The temporal order description provides better comparisons with experimental measurements in the time domain, which concerns the round-trip time (frequency) of each soliton rather than its detailed evolution within a single round trip. The temporal order description is sketched in Fig. S1. Resembling the experimental measurement, we divide the output signal in the time domain using a fixed frame of span $T_{\mathrm{R}} = 1/f_0 = 1/v_0$ in which $v_0$ is the corresponding averaged group velocity (assuming a unitary cavity length). Then the soliton within each frame acquires a relative frame time $\tau_k$ (for the $k^{\mathrm{th}}$-soliton). Then the incremental change of the frame time for the $k^{\mathrm{th}}$ soliton can be written as:

$$\delta \tau_k = T_{\mathrm{R}} \frac{d\tau_k}{dT} = -\frac{\Delta v_k}{v_0^2} + \Phi_k \tag{S2}$$

The minus signs on the RHS result from the space-time relation for a propagating signal. The term $-\Delta v_k/v_0^2$ also equals to $-\Delta f_k/f_0^2$, where $\Delta v_k$ stands for the group-velocity offset from $v_0$. $\Phi_k$ stands for a random step in time domain within the frame that also follows a normal distribution, i.e. $\Phi_k \sim \mathbf{N}(0, \sigma_\tau)$.

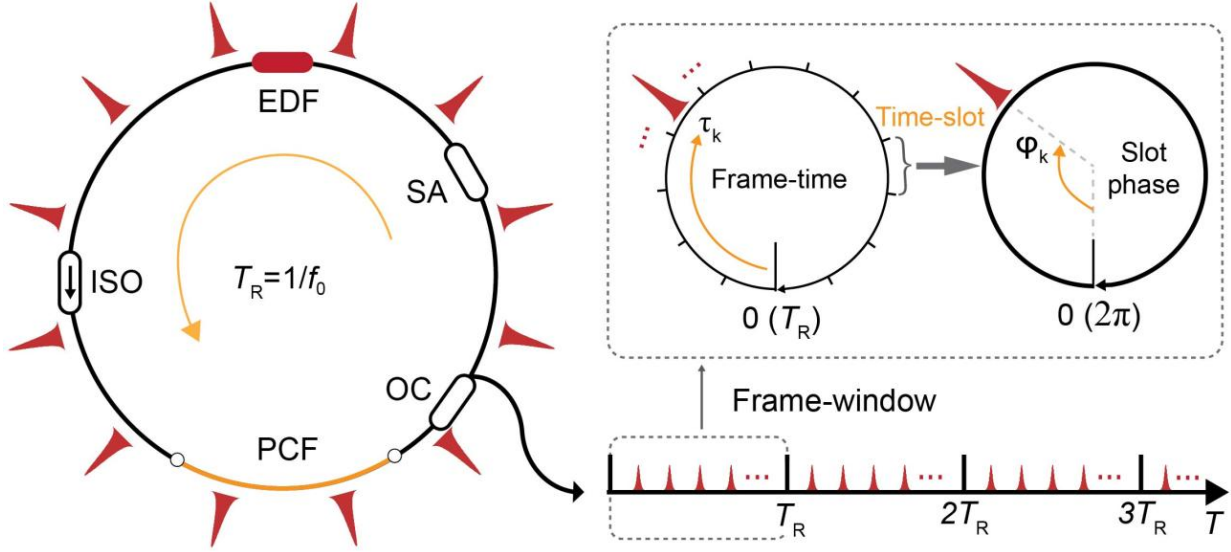

**Fig. S1 | Temporal order description of soliton ensemble in mode-locked lasers.** The harmonically mode-locked laser hosting an ensemble of $N_{\mathrm{p}}$ pulses can be described using the temporal order in the output sequence. The output of the laser is divided into a series of frames of span $T_{\mathrm{R}}$ which corresponds to a reference averaged group velocity $v_0$. Each frame is further divided into $N$ time-slots, each set to acquire an equivalent span of $2\pi$. The pulse located in the corresponding time-slot can then obtain a slot phase $\varphi_k$.

Within each frame, the frame time interval ($\Delta\tau_{kj}$) between two solitons ($k^{\mathrm{th}}$ and $j^{\mathrm{th}}$) is defined in a unidirectional way as if they are in a ring structure in the time domain as sketched in Fig.S2 below. Thus $\Delta\tau_{kj}$ is always positive for $k \neq j$. For $\tau_k - \tau_j > 0$, we have $\Delta\tau_{kj} = \tau_k - \tau_j$, while for $\tau_k - \tau_j < 0$, we have $\Delta\tau_{kj} = \tau_k - \tau_j + T_{\mathrm{R}}$. Each frame is further divided into $N$ time-slots according to the harmonic order $N$ of final steady state. A slot phase $\varphi_k$ can be obtained for each pulse located in its corresponding time-slot for calculating the order parameter $R$.

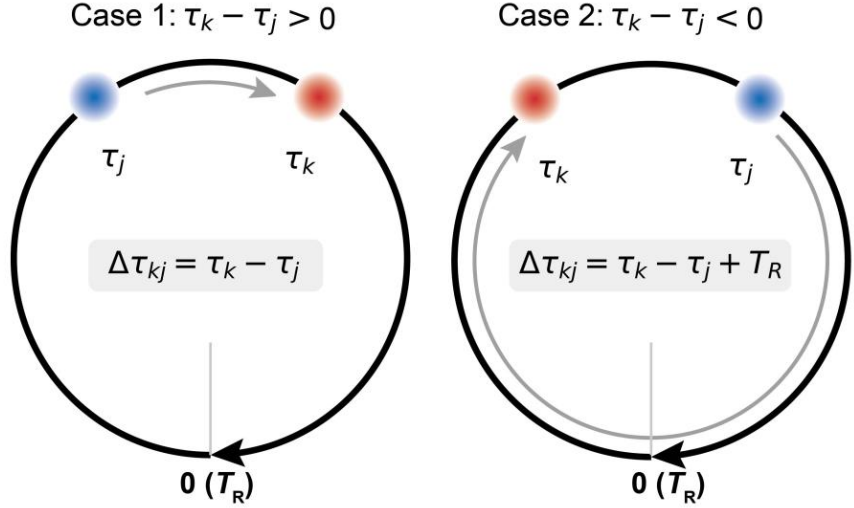

**Fig. S2 | Definition of the frame-time interval between two solitons**. The frame-time interval $\Delta\tau_{kj}$ between two solitons (with frame time at $\tau_k$ and $\tau_j$ respectively) on the ring is defined in a unidirectional way according to their temporal order. For $\tau_k - \tau_j > 0$, we have $\Delta\tau_{kj} = \tau_k - \tau_j$, while for $\tau_k - \tau_j < 0$, we have $\Delta\tau_{kj} = \tau_k - \tau_j + T_{\mathrm{R}}$.

In the model description, the dynamics of each particle on the ring is modelled by a two-dimensional system, which was then simplified to one-dimensional following the overdamped assumption. The two-dimensional system used the velocity offset $\Delta v_k$ from the reference group velocity $v_0$ of the frame and the corresponding frame time $\tau_k$ over consecutive frames as the two state variables for each particle flowing on the ring.

## 2. Details of experimental setups

The experimental setup that we build to study self-organization process is briefly sketched in Fig.S3, which consists of four parts: (i) the optoacoustically mode-locked fiber laser, (ii) the fast laser gain switch based on externally launched laser, (iii) the homodyne detection system for measuring the phase modulation in PCF, and (iv) the diagnostic for measuring temporal and spectral signal from the laser.

In the mode-locked laser cavity, the high-harmonic mode-locking (HHML) is achieved through the combined effects of nonlinear polarization rotation (NPR) and opto-acoustic interaction in the photonic crystal fiber (PCF). The total cavity length is ~ 34 m, with the laser gain provided by an ~ 1-m-long erbium-doped fiber (EDF) bidirectionally pumped by two 980-nm laser diodes. An ~ 0.7-m-long PCF with an ~1.9-μm-diameter core provides highly enhanced optoacoustic interaction, supporting an acoustic resonance at 1.905 GHz. In the steady state, the pulse repetition rate is locked at 1.90 GHz with the corresponding harmonic order of 310. In addition, a tunable delay-line (TD) enables the fine adjustment of the cavity round-trip frequency such that a specific harmonic lies in the acoustic resonance of the PCF and thereby allowing for precise control of the final repetition-rate and the harmonic order.

The fast laser-gain switch is enabled by depletion of the laser gain through externally launching a CW laser into the EDF section of the mode-locked cavity. The on/off switching of the external CW signal is controlled by a fast optical switch. By applying electric pulse using a pulse pattern generator, the external CW laser can transiently be launched into the EDF gain of the cavity through an 80/20 coupler and deplete the laser gain. Then after the external laser is switched off, the laser gain is restored, and the self-organization process will be kicked off.

The homodyne detection setup measures the accumulated phaser modulation in the PCF during the self-organization process through Mach-Zehnder-type interferometry. A single-frequency laser was split into two paths, one being a reference and the other being probe. The probe light was launched into the PCF, and the transmitted probe interferes with the reference part at the 50/50 coupler, which converts the acoustically induced phase modulation in the PCF into power modulation, which can be recorded by a photodetector and oscilloscope. To linearize the detection response, the interferometer was operated at the quadrature point. Both parts were amplified and adjusted using FPCs to maximize the interference contrast. The probe path inside the cavity was terminated with an in-line polarizer to avoid perturbing the EDF in the cavity.

The diagnostic setup detects the time domain signal using a fast oscilloscope (with an analog bandwidth of 59 GHz and a sampling rate of 200 G/s) and the spectral domain signal using an optical spectral analyzer (with a resolution of 0.02 nm).

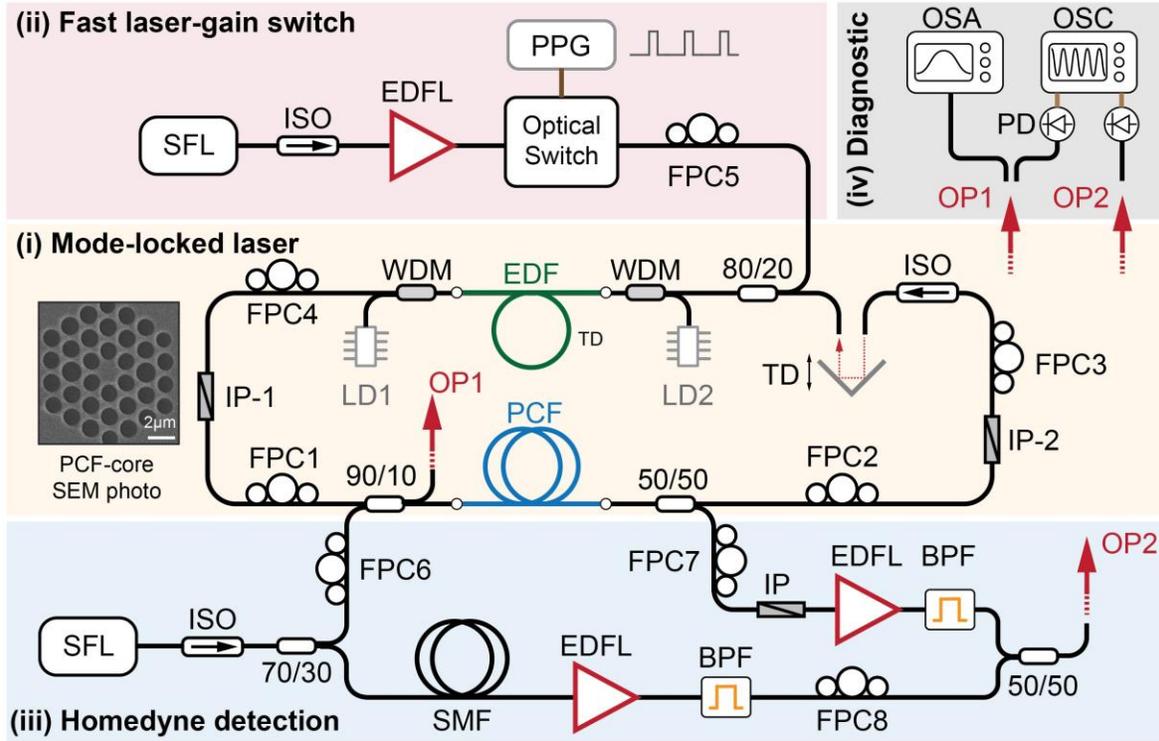

**Fig. S3 | Experimental setup. (i)** Mode-locked laser: FPC, fibre polarization controller; IP, in-line polarizer; WDM, wavelength division multiplexing coupler; LD, laser diode; EDF, erbium-doped fibre; TD, tuneable time-delay; ISO, isolator; PCF, photonic crystal fibre; **(ii)** Fast laser-gain switch: SFL, single frequency laser; EDFL, erbium-doped fibre laser; PPG, pulse pattern generator. **(iii)** Homedyne detection: BPF, band-pass filter. **(iv)** Diagnostic: OSA, optical spectrum analyser; PD, photo detector; OSC, oscilloscope.

## 3. Details about the low-dimensional dynamic model

### 3.1 Optoacoustic coupling function

The optoacoustic interactions induced changes to the velocity shift $\Delta v_k$, which entails a two-dimensional system that includes the accelerations $\delta \Delta v_k$. We have denoted the optomechanical coupling function using the function $F_A(\tau_k, \tau_j)$, such that it generates a discrete change of velocity upon the $k^{\text{th}}$ soliton as $\delta \Delta v_{k,\text{A}} = \sum_{j=1}^{N_\text{p}} F_A(\tau_k, \tau_j)$ as derived in the main text. In a rigid sense, since the acoustic resonance of the PCF has comparable lifetime with the cavity round-trip time in experiments, the impact of optoacoustic interactions with solitons in previous round-trips (including the same soliton itself) should also be included in this interaction function. If we denote the current round-trip number as $M$, then the frame time of the $j^{th}$ pulse in all previous round trips can be denoted as $\tau_{j,M-l}$, $l = 0, 1, 2, \ldots, (M-1)$. Then the total impact upon the $k^{th}$ soliton due to optoacoustic effect at current round-trip $M$ can be rewritten as

$$\delta\Delta v_{k,A} = \sum_{l=0}^{M-1}\sum_{j=1}^{N_P} F_A\big(\tau_{k,M}, \tau_{j,M-n}\big) \tag{S3}$$

The frame time difference between $\tau_{k,M}$ and $\tau_{j,M-n}$ in function $F_A$ can be expressed as

$$\Delta\tau_{kj,M-l} = \tau_{k,M} - \tau_{j,M-l} + lT_R \tag{S4}$$

In practice, the numerical simulation results suggest that only the impact from the last two round-trips needs to be included, while those from earlier ones can be neglected in the calculation.

### 3.2 Conditions for soliton collisions

The collision-induced coupling function $D_v$ and $D_\tau$ are non-zero only if the spacing between two solitons falls below a certain threshold value $\varepsilon_{th}$. As we observed in experiments, the outcomes of the collision are unsymmetric for the two solitons depending on their temporal order. Therefore, the condition for soliton collisions in theoretical analysis and numerical simulations needs further elaboration to make the modelling self-consistent. In addition, the simulation also needs to deal with the non-trivial cases with multiple narrowly-spaced solitons, such that the soliton sandwiched in the middle needs to have consistent rules for collisions.

If we consider a specific soliton, e.g. the $k^{th}$ soliton at the frame time of $\tau_k$, the other soliton that collides with it (as denoted as the $j^{th}$ soliton at a frame time of $\tau_{j,C}$) may be ahead of or behind it. For the $k^{th}$ soliton that collides with a soliton ahead of it, their frame time interval fulfills the following conditions:

#### __Conditon I__

$$\Delta\tau_{kj,C} = \mathbf{min}\{\Delta\tau_{kj}\} < \varepsilon_{th} \quad \& \ \Delta\tau_{kj,C} \le \mathbf{min}\{T_R - \Delta\tau_{kj}\} \text{ for } j \ne k \tag{S5}$$

The second part of the condition (RHS) in (S5) deals with the case in which the $k^{th}$ soliton is intimately sandwiched by two other solitons, the soliton in the middle is set to first collide with its nearest neighbor, which is the one ahead of it.

Meanwhile, for the $k^{th}$ soliton to collide with a soliton lagged it, the condition reads:

<u>**Condition II**</u>

$$T_R - \Delta\tau_{kj,C} = \mathbf{min}\{T_R - \Delta\tau_{kj}\} < \varepsilon_{th} \ \& \ T_R - \Delta\tau_{kj,C} < \mathbf{min}\{\Delta\tau_{kj}\} \text{ for } j \neq k$$

Similarly, the second part of the condition (RHS) in (S6) deals with sandwiched structure. For two intimately spaced solitons, Conditions I and II will be fulfilled for the two colliding solitons, respectively. Then the different outcomes can be specified for the two solitons, respectively. From the perspective of the $k^{th}$ soliton, the collision-induced coupling function would be Eqs. (4) and (5) for those fulfilling Condition I and Eqs. (6) and (7) for those fulfilling Condition II as given in Methods. In the rare case of three equally and intimately spaced solitons, the soliton in the middle is set to first collide with the one ahead of it in time. These two conditions for soliton collisions are sketched in Fig.S4.

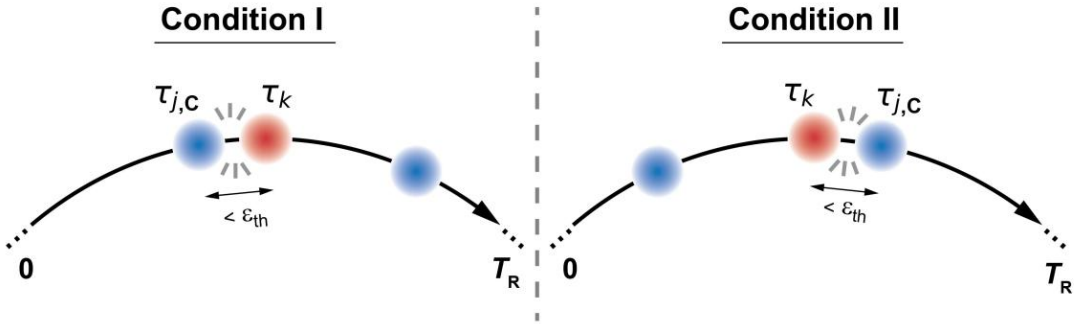

**Fig. S4 | Conditions for soliton collisions.** In Condition I, the soliton collides with the soliton intimately ahead of it with spacing below threshold $\varepsilon_{th}$. In Condition II, the soliton collides with soliton intimately behind it with spacing below threshold $\varepsilon_{th}$. In both cases, the soliton sandwiched by two other solitons is set to collide with the one being closer to it.

## 3.3 Resemblance with Kuramoto model

The low-dimensional model described in Eq.(5) in the main text resembles the well-known Kuramoto model that can be used to describe the self-synchronization process of coupled oscillators. A typical Kuramoto model for $N$-coupled oscillators with all-to-all coupling topology can be described by $N$-coupled equations as below:

$$\frac{d\varphi_k}{dT} = \omega_k + \frac{\varepsilon}{N}\sum_{j=1}^{N}F(\varphi_k, \varphi_j), k = 1,2,\dots N \tag{S9}$$

in which $\omega_k$ represent the intrinsic frequency of each oscillator, which are typically different from each other and $\varepsilon$ represent the coupling strength between the oscillators. The function $F(\varphi_k, \varphi_j)$ represent the nonlinear interactions between each pair of coupled oscillators, which usually takes a sinusoidal dependence upon their phase difference, i.e.

$$F(\varphi_k, \varphi_j) = \sin(\varphi_j - \varphi_k) \tag{S10}$$

In our case, the long-range interaction function $F_A(\tau_k, \tau_j)$ in has a real part that reads

$$\mathbf{Re}\{F_A(\tau_k, \tau_j)\} = \frac{C_0 A}{\gamma} e^{-\Gamma_q \Delta \tau_{kj}} \sin\left(\sqrt{\Omega_q^2 - \Gamma_q^2} \, \Delta \tau_{kj}\right) \tag{S11}$$

Therefore, we can readily see both Eq.(4) in the main text and Eq.(S9) above describe a one-dimensional equation for $N$-oscillator system with sinusoidal coupling between them depending on the "phase difference" between these oscillators. Therefore, it is not surprising that the current mode-locked laser system can also exhibit self-organization process as self-synchronized coupled-oscillators, since they follow similar mathematical systems.

Meanwhile, some important difference exists between these two interaction terms. Firstly, in our case, the interaction is rendered by an exponential decay that depends on the frame time gap. This decay results from the acoustic damping effect that corresponds to the bandwidth of the acoustic resonance. The non-trivial acoustic resonance allows for a final soliton repetition rate ($Nf_0$) that slightly deviates from the resonance frequency ($\Omega_q$). Secondly, all the solitons are set to have an identical intrinsic repetition rate $f_0$ while suffering perturbation from noise ($\Phi_{k,\tau}$) which demands interactions of sufficient strength to overcome. At last, our system acquires a unique short-range collision-type interaction that does not appear in the classical Kuramoto model. However, the existence of collisions only causes minor changes in the detailed dynamics and does not affect the general trend of self-organization of the soliton ensemble, as we observed in both experiments and simulation results.

## 4. Different possibilities of soliton collisions

Typical outcomes of the soliton collisions, as we observed in the self-organization process in experiments, can be categorized into two possibilities. The first possibility is that one of the solitons is eliminated while the other one survives and continues to propagate. In the main text, we have mainly discussed this possibility. However, a second possibility exists such that the two solitons strongly repel each other and gain a non-trivial spacing between them shortly after the collision. However, due to the optoacoustic coupling, the repelled solitons will fall into collision range later and trigger the collision-induced actions again. Such possibility will typically lead to consecutive collisions of two solitons even though the basic temporal pattern (and the optomechanical lattice) has been established.

Like the Eqs. (4) – (7), we also set coupling function in this possibility as a simple lumped-action function over a single iteration following the rules below. Firstly, the two solitons will exchange their current velocity (such that their spacing will diverge instead of further converge). Secondly, we can set that the soliton ahead in time (i.e. **Condition II** is fulfilled) experienced a random advancement in its frame time from its current value by a limited amount while the other one remains in its previous position. Then we can formulate that for solitons that fulfill either **Condition I or II**, we have

$$D_v\big(\Delta v_k, \Delta v_{j,C}\big) = -\Delta v_k + \Delta v_{j,C}, \tag{S7}$$

in which $\Delta v_{j,C}$ is the current velocity offset of the soliton that collides with the $k^{\text{th}}$ soliton. Meanwhile, for solitons that fulfill **Condition I**, we set $D_\tau\big(\tau_k, \tau_{j,C}\big) = 0$, for solitons that fulfill **Condition II**, we set random advancement in the frame time to follow a uniform distribution within a narrow window (half of the averaged spacing between neighboring solitons in the cavity), i.e.

$$D_\tau\big(\tau_k, \tau_{j,C}\big) = -\eta_k, \qquad \eta_k \sim \mathbf{U}(0, T_{\text{R}}/2N) \tag{S8}$$

The simplified rules for the collision-induced coupling functions in these two possibilities are summarized in Fig.S5.

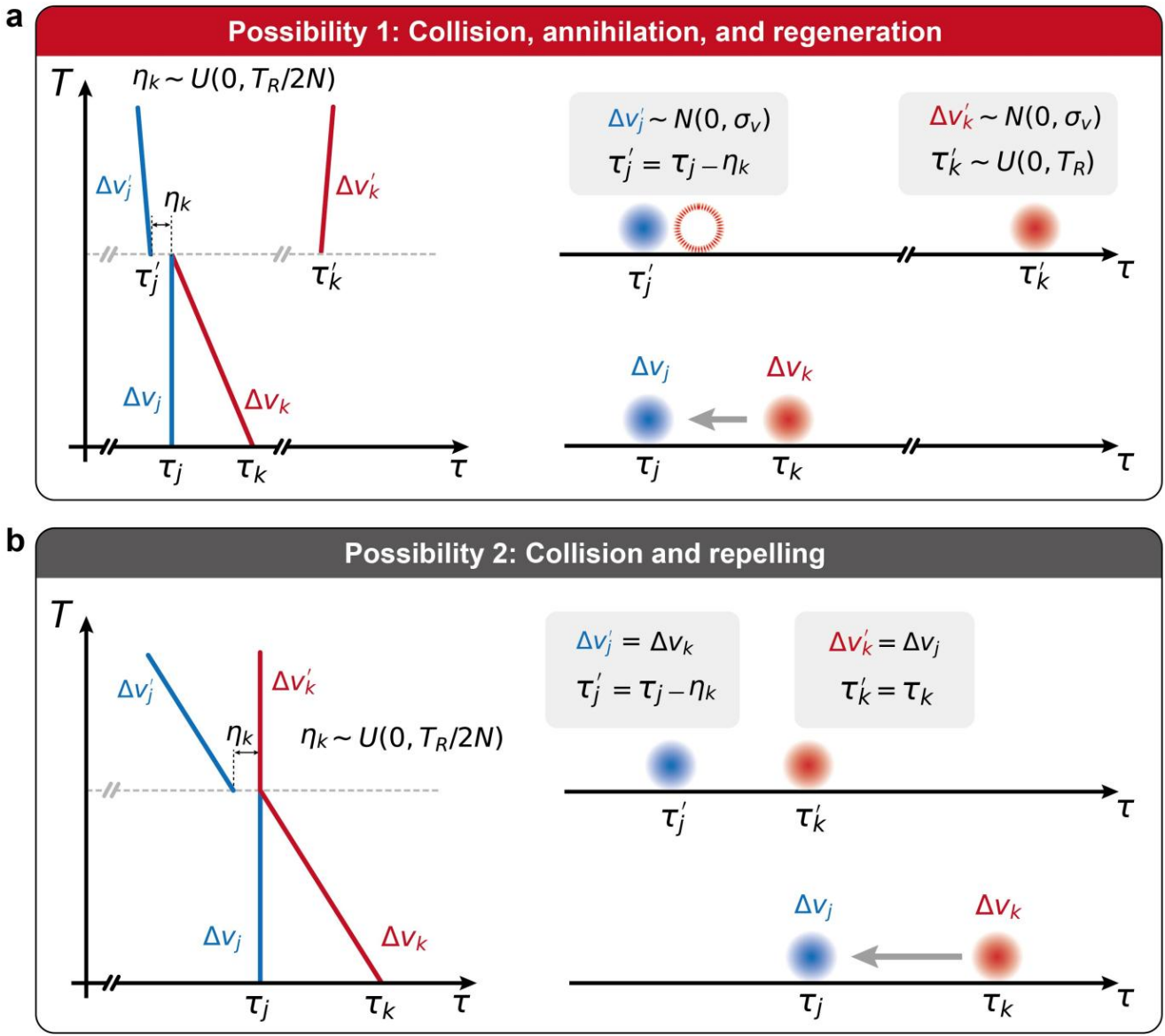

**Fig. S5 | Simplified rules for two possibilities of collision-induced coupling**. **a.** Possibility I: Collison-annihilation-regeneration. When the two solitons collide, the earlier one remains in position with averaged velocity, the other one is annihilated and regenerated at a random position with a random velocity. **b.** Possibility II: Collision-repelling. The colliding solitons exchange their velocities, while the earlier is pushed by a non-trivial distance and the other one remains in the same position.

In experiments, we have observed the types of collisions under different cavity parameters. Typical recordings are shown in Fig.S6. In some cases, the two colliding solitons quickly ended up with one soliton (Fig.S6a), while in some other cases, the two colliding solitons strongly repel each other after the collisions, while being trapped, leading to consecutive collisions lasting for tens to hundreds of ms. Such type of collision, once occurred, will typically delay the stabilization of the entire pattern by a significant extent in the self-organization process. The exact mechanism for these different features of collision dynamics remains to be explored, while the basic feature of the outcomes can be essentially captured using simplified coupling functions in the low-dimensional model.

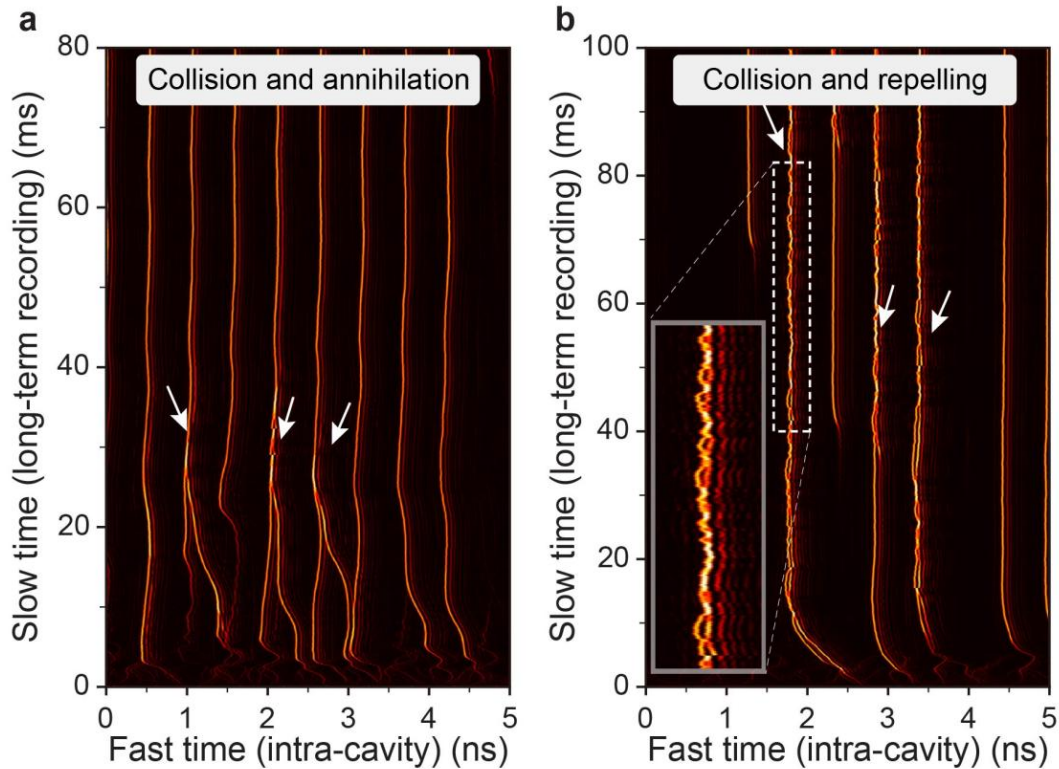

**Fig. S6 | Two possibilities of soliton collisions observed in self-organization process**. **a**. Collison-annihilation processes. The two colliding solitons quickly ended up with only one soliton within a few ms. **b**. Collision-repelling process. The two colliding solitons repel each other strongly after collision while being trapped, leading to consecutive collisions lasting for tens to hundreds of ms.